\documentclass{ws-mpla}
\usepackage{bbm}
\usepackage{amsfonts}
\begin{document}

\markboth{X.-H. Zhai, X.-Z. Li and C.-J. Feng} {Casimir effect with
a helix torus boundary condition}

\catchline{}{}{}{}{}

\title{Casimir effect with a helix torus boundary condition}

\author{\footnotesize XIANG-HUA ZHAI\footnote{zhaixh@shnu.edu.cn}, XIN-ZHOU LI\footnote{kychz@shnu.edu.cn} and
CHAO-JUN FENG\footnote{fengcj@shnu.edu.cn} }

\address{Shanghai United Center for Astrophysics (SUCA), Shanghai Normal University,
    100 Guilin Road, 100 Guilin Road, Shanghai 200234, China
    }

\maketitle

\pub{Received (Day Month Year)}{Revised (Day Month Year)}

\begin{abstract}
We use the generalized Chowla-Selberg formula to consider the
Casimir effect of a scalar field with a helix torus boundary
condition in the flat ($D+1$)-dimensional spacetime.
 We obtain the exact results of
the Casimir energy density and pressure for any $D$ for both
massless and massive scalar fields. The numerical calculation
indicates that once the topology of spacetime is fixed, the ratio of the sizes of
the helix will be a decisive factor. There is a critical value
$r_{crit}$ of the ratio $r$ of the lengths at which the pressure
vanishes. The pressure changes from negative to positive as the
ratio $r$ passes through $r_{crit}$ increasingly. In the massive
case, we find the pressure tends to the result of massless field
when the mass approaches zero. Furthermore, there is another
critical ratio of the lengths $r_{crit}^{\prime}$ and the pressure
is independent of the mass at $r=r_{crit}^{\prime}$ in the $D=3$
case.

\keywords{Casimir effect; zeta function; boundary condition;
Chowla-Selberg formula.}
\end{abstract}

\ccode{PACS Nos.: 02.30.Gp; 11.10.-z}

\section{Introduction}

Casimir's calculation of the force between two neutral, parallel
conducting plates originally inspired much theoretical interest as
macroscopic manifestation of quantum fluctuation of the field in
vacuum. However, the Casimir effect arises not only in the presence
of material boundaries, but also in spaces with non-Euclidean
topology\cite{1}. The simplest example of the Casimir effect of
topological origin is the scalar field on a flat manifold with
topology of a circle $\textit{S}^1$. The topology of $\textit{S}^1$
causes the periodicity condition $\phi(t,0)=\phi(t,C)$ for a
Hermitian scale field $\phi(t,x)$, where $C$ is the circumference of
$\textit{S}^1$, imposed on the wave function which is of the same
kind as those due to boundary and resulting in an attractive Casimir
force. Similarly, the antiperiodic conditions can be drawn on a
M\"{o}bius strip and bring about the repulsive Casimir force as a
result. Recently, the topology of the helix boundary conditions is
investigated in ref.\cite{2}. We find that the Casimir effect is
very much like the effect on a spring that obeys the Hooke's law in
mechanics. However, in this case, the pressure comes from a quantum
effect, so we would like to call this structure a quantum spring.
The pressurre is negative in both massless and massive scalar cases
for this structure\cite{3}.

It is worth noting that the concept of quotient topology is very
useful for concrete application. We consider a surjective mapping
$f$ from a topological space $X$ onto a set $Y$. The quotient
topology on $Y$ with respect to $f$ is given in \cite{4}. Surjective
mapping can be easily obtained when we use the equivalence classes
of some equivalence relation $\sim$. Thus, we let $X/\sim$ denote
the set of equivalence classes and define $f: X\rightarrow X/\sim$
by $f(x)=[x]$ the equivalence class containing $x$. $X/\sim$ with
the quotient topology is called to be obtained from $X$ by
topological identification. For example, if we take the rectangle
$Y=\{(x^1, x^2); 0\leq x^1 \leq a,0 \leq x^2 \leq h\}$ in
$\mathbb{R}^2$ with the induced topology and define an equivalence
relation $\sim$ on $Y$ by $(x^1,x^2)\sim(x^{1\prime},
x^{2\prime})\Leftrightarrow (x^1,x^2)=(x^{\prime1}, x^{\prime2})$ or
$ \{x^1, x^{\prime1}\}=\{0,a\}$ and $x^2=x^{\prime2}$, then $Y/\sim$
with the quotient topology is homomorphic to the cylinder
$C=\{(x,y,z)\in \mathbb{R}^3; x^2+y^2=\left(\frac a{2\pi}\right)^2,
|z|\leq 1\}.$ The boundary condition $\phi(t,0,x^2)=\phi(t,a,x^2)$
can be drawn on the topology of a cylinder. Similarly, we define
another equivalence relation $\sim$ on $Y$ by
$(x^1,x^2)\sim(x^{\prime1}, x^{\prime2})\Leftrightarrow
(x^1,x^2)=(x^{\prime1}, x^{\prime2})$ or $ x^1=0, x^{\prime1}=a,
x^2=x^{\prime2}$ and $x^1=x^{\prime1}, x^2=0, x^{\prime2}=h$, then
$Y/\sim$ with the quotient topology is homomorphic to a torus
$T=\{(x,y,z)\in \mathbb{R}^3; \left(\frac
a{2\pi}-\sqrt{x^2+y^2}\right)^2+z^2=\left(\frac h{2\pi}\right)^2, 0
\leq h \leq a\}$. The boundary conditions
$\phi(t,0,x^2)=\phi(t,a,x^2)$ and $\phi(t,x^1,0)=\phi(t,x^1,h)$can
be drawn on the topology of the torus. In this paper, we will
consider the helix torus topology using the concept of quotient
topology.

The $\zeta$-function regularization procedure is a powerful and
elegant technique for the Casimir effect \cite{5}. The generalized
$\zeta$-function has many interesting applications, e.g., in the
piecewise string \cite{6,7}. Similar analysis has been applied to
rectangular cavity \cite{8}-\cite{10}, noncommutative
spacetime\cite{11}, p-branes \cite{12} or pistons
\cite{13}-\cite{17}. Casimir effect for a fractional boundary
condition is of interest in considering, for example, the finite
temperature Casimir effect for a scalar field with fractional
Neumann conditions \cite{18}, while the repulsive force from
fractional boundary conditions has been studied \cite{19}. The
Chowla-Selberg formula of the $\zeta$-function has been applied to
quantize the Wheelar-DeWitt equation \cite{20} and recently to calculate the Casimir energies of cylinders \cite{21}.

In this paper, we consider the Casimir effect of a scalar field with
a helix torus boundary condition in the flat ($D+1$)-dimensional
spacetime. The Chowla-Selberg formula and its generalization are
used to regularize the Casimir energy density. We obtain the exact
results of the Casimir energy and pressure for any $D$ for both
massless and massive scalar fields. The numerical calculation
indicates that once the topology of spacetime is fixed, the ratio of the sizes of
the helix will be a decisive factor. In massless case, there is a
critical value $r_{crit}$ of the ratio $r$ of the lengths at which
the pressure vanishes. The pressure changes from negative to
positive as the ratio $r$ passes through $r_{crit}$ increasingly. In
massive case, we compare the pressure of massive field with massless
one and find the pressure tends to the result of massless field when
the mass approaches zero. Furthermore, we find there is also a
critical ratio $r_{crit}(\mu)$ at which the pressure vanishes. At
the same time, there is another critical ratio of the lengths
$r_{crit}^{\prime}$ and the pressure is independent of the mass at
$r=r_{crit}^{\prime}$ in the $D=3$ case. The outline of this paper
is as follows. In Sec. 2 we introduce a specified helix topology for
the flat spacetime. In Sec. 3 we consider the evaluation of the
Casimir effect for massless and massive cases. The results are
summarized in Sec. 4.

\section{Helix torus topology}

The Casimir effect arises not only in the presence of material
boundaries, but also in spaces with non-Euclidean topology. First,
we consider a specified 2-dimensional helix topology which may be
obtained by identifying some or whole in a series of rectangles as a
pedagogical discussion. The whole situation
 concerning the helix boundary conditions in this and similar cases is discussed
 in \cite{2,3} in great detail.  In Fig. 1(a), the cylinder is obtained by identifying
  some of the boundary points of a rectangle, in which we intend to identify the two edges.
  This is often indicated by labeling the two points of edges with the same letter, such as A, B.
   Similarly, we may obtain a torus (see Fig. 1(b)). Furthermore, a
   helix torus topology can occur by identifying both distinct
   rectangles and the boundary points at the same rectangle(see Fig.
   1(c)).
\begin{figure}[ph]
\centerline{\psfig{file=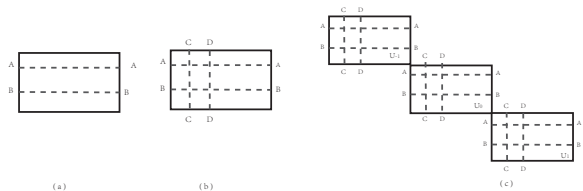,width=3.0in}}
\vspace*{8pt}
\caption{There are many topologies that may be obtained by identifying various boundary points
of a rectangle or many rectangles. (a) A cylindrical topology: we may equally well picture the
 cylinder as being the topological space obtained by identifying the two edges. This is indicated by
  labeling the two points of edges with the same letter. (b) A torus topology: we may equally well
  picture the torus as being the topological space obtained by identifying the two pairs of edges in a rectangle.
  In other words, a torus topology occurs if we join two side of
  rectangle, then join the other perpendicular direction as well.
  (c) A helix torus topology:
  A helix topology may be obtained by identifying some or whole in a series of rectangles,
  in contrast to the case of cylindroid or torus. A helix torus topology may be obtained by
   identifying both distinct rectangles and the boundary points at the same rectangle.
   In this figure, we show 3 rectangles which are denoted $U_i$ ($i=-1,0,1$) respectively.
   Here, $U_{-1}=\{x^1\mathbbm{e}_1+x^2\mathbbm{e}_2|-a\leq x^1\leq0,h\leq x^2\leq2h\},
   U_0=\{x^1\mathbbm{e}_1+x^2\mathbbm{e}_2|0\leq x^1\leq a,0\leq x^2\leq h\}$ and
   $U_1=\{x^1\mathbbm{e}_1+x^2\mathbbm{e}_2|a\leq x^1\leq 2a,-h\leq x^2\leq 0\}$, and every $U_i$ has torus topology.}
\end{figure}

Before we consider a helix torus topology in the flat
($D$+1)-dimensional spacetime $\mathcal{M}^{D+1}$ using the concept of quotient topology, we have to
discuss the lattices. A lattice $\Lambda$ is defined as a set of
points
\begin{equation}\label{eq1}
    \Lambda = \left\{ ~ \sum_{i=0}^{D} n_i \mathbbm{e}_i ~|~ n_i \in \mathcal{Z} ~\right\} \,,
\end{equation}
where $\{\mathbbm{e}_i\}$ is a set of basis vectors of
$\mathcal{M}^{D+1}$. In terms of the components $v^i$ of vectors
$\mathbb{V} \in \mathcal{M}^{D+1} $, we define the inner products as
\begin{equation}\label{eq2}
    \mathbb{V} \cdot \mathbb{W} = \epsilon(a)v^iw^j\delta_{ij} \,,
\end{equation}
with $\epsilon(a)=1$ for $i=0$, $\epsilon(a)=-1$ for otherwise. In the $x^1-x^2$ plane, the sublattice
$\Lambda''\subset\Lambda'\subset\Lambda$ are
\begin{equation}\label{eq3}
   \Lambda' = \left\{ ~  n_3 \mathbbm{e}_1 + n_2 \mathbbm{e}_2 ~|~ n_2,n_3 \in \mathcal{Z} ~\right\} \,,
\end{equation}
and
\begin{equation}\label{eq4}
   \Lambda'' = \left\{ ~  n_1(\mathbbm{e}_1 - \mathbbm{e}_2) ~|~ n_1 \in \mathcal{Z} ~\right\} \,.
\end{equation}
The unit cell is the set of points
\begin{eqnarray}\label{eq5}
   C_0 = \bigg\{\mathbf{X} &=& \sum_{i=0}^{D}x^i \mathbbm{e}_i ~|~ 0\leq x^1 \leq a,
 -h\leq x^2 \leq 0 ,  (0,x^2)\Leftrightarrow(a,x^2) \nonumber\\
   & &\mathrm{and} (x^1,0)\Leftrightarrow(x^1,-h);-\infty <x^0<\infty, -\frac{L}{2} \leq x^T\leq \frac{L}{2}\bigg\} \,,
\end{eqnarray}
where $T = 3,\cdots, D$ and $\Leftrightarrow$ is a symbol of
identity relation.

Next, we choose topological space $\mathbf{X}$ as
$\mathbf{X}=\bigcup_{\mathbbm{u}\in \Lambda^{''}
}\{C_0+\mathbbm{u}\}$ in $\mathcal{M}^{D+1}$ with the induced
topology and define an equivalence relation $\sim$ on $\mathbf{X}$
by $(x^1,x^2)\sim(x^1-a,x^2+h)$, then $\mathbf{X}/\sim$ with the
quotient topology is a new type topology which can be called a helix
torus topology. This topology causes the helix boundary condition
for a Hermitian scalar field
\begin{equation}\label{eq6}
   \phi(t, x^1 + a, x^2, x^T) =  \phi(t, x^1 , x^2+h, x^T) \,,
\end{equation}
and the periodicity boundary conditions
\begin{equation}\label{eq7}
   \phi(t, x^1 , 0, x^T) =  \phi(t, x^1 , h, x^T) \,,
\end{equation}
and
\begin{equation}\label{eq8}
   \phi(t, -a , x^2, x^T) =  \phi(t, 0 , x^2, x^T).
\end{equation}
It is worth noting that Eq.(\ref{eq8}) can be derived from
Eqs.(\ref{eq6}) and (\ref{eq7}), or  Eq.(\ref{eq7}) derived from
Eqs.(\ref{eq6}) and (\ref{eq8}). Therefore, we need only one of the
two periodicity  boundary conditions.

\section{Evaluation of the Casimir effect}

\subsection{The Casimir energy density}

In calculations on the Casimir effect, extensive use is made of
eigenfunctions and eigenvalues of the corresponding field equation.
A Hermitian scalar field $\phi(t, x^\alpha, x^T)$ defined in the
($D+1$)-dimensional flat spacetime satisfies the Klein-Gordon
equation:
\begin{equation}\label{eq9}
    \left(\partial_t^2 - \partial_i^2+\mu^2\right)\phi(t, x^\alpha, x^T) = 0 \,,
\end{equation}
where $i=1,\cdots, D; \alpha=1,2; T=3,\cdots, D$ and $\mu$ is the
mass of the scalar field. Under the boundary condition (\ref{eq6})
and (\ref{eq7}), the modes of the field are then
\begin{equation}\label{eq10}
    \phi_{n}(t, x^\alpha, x^T)= \mathcal{N} e^{-i\omega_nt+ik_x x+ik_z z + ik_Tx^T }\,,
\end{equation}
where $\mathcal{N}$ is a normalization factor and $x^1=x, x^2=z$, and we have
\begin{equation}\label{eq11}
    \omega_n^2 = k_{T}^2 + k_x^2 + \left( -\frac{2\pi n_1}{h}+\frac{k_x}{h}a \right)^2+\mu^2 =
     k_{T}^2 + k_z^2 + \left( \frac{2\pi n_1}{a}+\frac{k_z}{a}h
    \right)^2+\mu^2 \,.
\end{equation}
Here, $k_x$ and $k_z$ satisfy
\begin{equation}\label{eq12}
    a k_x - hk_z = 2n_1\pi,
\end{equation}
\begin{equation}\label{eq13}
k_z=\frac{2\pi n_2}{h},k_x=\frac{2\pi n_3}{a},
\end{equation}
where $n_2,n_3=0,\pm1,\pm2,\cdots$ and the constraint condition
$n_1=n_3-n_2$. In the ground state (vacuum), each of these modes
contributes an energy of $\omega_n/2$. The energy density of the field is
thus given by

\begin{eqnarray}\label{eq14}
\nonumber
  &\varepsilon^{D}& = \frac{1}{2a}
  \int \frac{d^{D-2}k_T}{(2\pi)^{D-2}} \sum_{n_1,n_2=-\infty}^{\infty \prime} \sqrt{k_T^2 + \left(\frac{2\pi n_2}{h}\right)^2+\left( \frac{2\pi n_1}{a}
  +\frac{2\pi n_2}{a} \right)^2 +\mu^2 } \,, \\&&
\end{eqnarray}

\noindent where we have assumed $a\neq 0$ and  $h\neq 0$ without
losing generalities and the prime on the summation means that
$(n_1,n_2)=(0,0)$ have to be omitted. Remember that there are no
material boundaries here and the Casimir effect arises owing to the
nontrivial topology. This allows us to define not only the
separation-dependent global energy but also the energy density as
above.

Using the mathematical identity
\begin{equation}
\int_{-\infty}^{\infty}f(u)d^{D-2}u=\frac{2\pi^{\frac{D-2}2}}{\Gamma\left(\frac{D-2}2\right)}\int_0^{\infty}u^{D-3}f(u)du,
\label{eq15}
\end{equation}
Eq.(\ref{eq14}) can be reduced to
\begin{eqnarray}
\varepsilon^D=&-&\frac 1 {2a} \pi^{\frac{D-1}2}\Gamma\left(-\frac {D-1}2\right)\nonumber\\
&\times&\sum_{n_1,n_2=-\infty}^{\infty
\prime}\left[\frac{n_1^2}{a^2}+\frac{2n_1n_2}{a^2}+\left(\frac
1{a^2}+\frac
1{h^2}\right)n_2^2+\left(\frac{\mu}{2\pi}\right)^2\right]^{\frac{D-1}2}.\label{eq16}
\end{eqnarray}

On the other hand, the generalized Chowla-Selberg formula was given
firstly by Elizalde et. al. \cite{5} and was derived in detail
recently by Abalo et. al.\cite{21}
\begin{eqnarray}
S&=&\Gamma(s)\sum_{m,n=-\infty}^{\infty \prime}(\alpha m^2+\beta
m n+\gamma n^2+\delta)^{-s}\nonumber\\
&=&2\Gamma(s)\zeta(s,\frac{\delta}{\alpha})\alpha^{-s}+\frac{2^{2s}\sqrt{\pi}\alpha^{s-1}}{\Delta^{s-\frac
{1}{2}}}\Gamma\Big(s-\frac{1}{2}\Big)\zeta\Big(s-\frac{1}{2},\frac{4\alpha
\delta}{\Delta}\Big)\nonumber\\
&+&4(2\pi)^s\sqrt{\frac{2}{\alpha}}\sum_{n=1}^{\infty}n^{s-\frac
{1}{2}}\cos(\frac{n\pi
\beta}{\alpha})\sum_{d|n}d^{1-2s}\Big(\Delta+\frac{4\alpha
\delta}{d^2}\Big)^{\frac{1-2s}{4}}\nonumber\\
&\times&K_{s-\frac{1}{2}}\Big(\frac{\pi
n}{\alpha}\sqrt{\Delta+\frac{4\alpha \delta}{d^2}}\Big).
\label{eq17}
\end{eqnarray}
where $\Delta=4\alpha\gamma-\beta^2$, $d|n$ denotes the divisor of $n$, $k_{\nu}(z)$ is the modified Bessel function and $\zeta(s,p)$ is the
Epstein-Hurwitz $\zeta$ function defined as
\begin{equation}
\zeta(s,p)\equiv\sum_{n=1}^{\infty}(n^2+p)^{-s}. \label{eq18}
\end{equation}
By using Eq.(\ref{eq17}), the infinite summation in Eq.(\ref{eq16})
can be regularized. In the following subsections, We discuss the
massless and massive cases respectively.
\subsection{The massless case}

In the massless case $\mu=0$, using Eq. (\ref{eq17}) with $\delta=0$
and the functional relation
\begin{equation}
\pi^{-\frac s2}\Gamma\Big(\frac
s2\Big)\zeta(s)=\pi^{-\frac{1-s}2}\Gamma\Big(\frac{1-s}2\Big)\zeta(1-s),
\label{eq19}
\end{equation}
\noindent  one can rewrite the
Casimir energy density of massless scalar field as
\begin{eqnarray}
\varepsilon^D_{\mu=0}=&-&\frac{\Gamma\left(\frac D 2\right)\zeta(D)}{\pi^{\frac D 2}a^D}-\frac{\Gamma\left(\frac{D+1}2\right)\zeta(D+1)}{\pi^{\frac{D+1}2}h^D}\nonumber\\
&-&\frac{4}{(ah)^{\frac D 2}}\sum_{n_1,n_2=1}^{\infty}\left(\frac{n_2}{n_1}\right)^{\frac D 2}K_{\frac D 2}\left(\frac{2\pi n_1n_2a}{h}\right),
\label{eq20}
\end{eqnarray}
where
\begin{equation}
\sum_{n_1n_2=1}^{\infty}\left(\frac{n_1}{n_2}\right)^{s-\frac
12}=\sum_{n=1}^{\infty}n^{s-\frac 12}\sum_{d|n}d^{1-2s}.\label{eq21}
\end{equation}

In the $D=3$ case, Eq.(\ref{eq20}) is reduced to
\begin{equation}
\varepsilon_{\mu=0}=-\frac{\zeta(3)}{2\pi a^3}-\frac{\pi^2
}{90h^3}-\frac{4}{a^{\frac 32}h^{\frac
32}}\sum_{n_1,n_2=1}^{\infty} \left(\frac{n_2}{n_1}\right)^{\frac 3
2}K_{\frac 3 2}\left(\frac{2\pi n_1n_2a}{h}\right). \label{eq22}
\end{equation}
From the thermodynamic relation, one can get the pressure on the
$x^1$ direction as
\begin{eqnarray}
P_{a,\mu=0}&=&-\frac{\partial E_{\mu=0}}{\partial a}\nonumber\\
&=&-\frac{\zeta(3)}{\pi a^3}+\frac{\pi^2}{90h^3}+\frac{2}{(ah)^{\frac 32}}\sum_{n_1,n_2=1}^{\infty}\left(\frac{n_2}{n_1}\right)^{\frac 3 2}K_{\frac 3 2}\left(\frac{2\pi n_1n_2a}{h}\right)\nonumber\\
&-&\frac{8\pi}{a^{\frac 12}h^{\frac 52}}\sum_{n_1,n_2=1}^{\infty}\left(\frac{n_2}{n_1}\right)^{\frac 1 2}n_2^2 K_{\frac 5 2}\left(\frac{2\pi n_1n_2a}{h}\right).
\label{eq23}
\end{eqnarray}
where $E_{\mu=0}=a\varepsilon_{\mu=0}$. It is worth noting that the pressure can also be obtained from the component of the energy-momentum tensor which has the physical meaning of a pressure.

By the numerical calculation, we have $P_{a,\mu=0}=0$ when
$a=r_{crit}h$ and $r_{crit}=1.52007606\cdots$. It is obvious that
$P_{a,\mu=0}$ is negative if $a<r_{crit}h$ and is positive if
$a>r_{crit}h$ since the first and fourth terms are negative and the
second and third terms are positive in Eq.(\ref{eq23}). When $a\gg
h$, we have $F_a\approx \frac{\pi^2}{90h^3}$. Fig. 2 is the
illustration of the behavior of the pressure on $x^1$ direction for
$D=3$. The curves correspond to $h=1,2,3,4$ respectively. It is
clearly seen that for a given $h$, the negative pressure become
positive with $a$ increasing. Once topology of spacetime is fixed,
the ratio of the sizes of the helix will be a decisive factor.

\begin{figure}[ph]
\centerline{\psfig{file=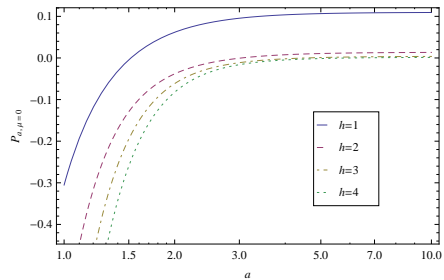,width=2.5in}} \caption{The pressure
$P_{a,\mu=0}$ on the $x^1$ direction for $D=3$ and $h=1,2,3,4$. It
is clearly seen that the pressure $P_{a,\mu=0}$ is a monotonic
function and vanishes at $a=r_{crit}h$ . It changes from attractive
to repulsive when passing through the critical ratio $r_{crit}$
increasingly and tends to $\frac{\pi^2}{90h^3}$. }
\end{figure}

\subsection{The massive case}

In the $\mu\neq 0$ case, using Eq. (\ref{eq17}), we have
\begin{eqnarray}
& &\Gamma\left(-\frac{D+1}{2}\right)\sum_{n_1,n_2=-\infty}^{\infty \prime}\left[\frac{n_1^2}{a^2}+\frac{2n_1n_2}{a^2}+\left(\frac {1}{a^2}+\frac{1}{h^2}\right)n_2^2+\left(\frac{\mu}{2\pi}\right)^2\right]^{\frac{D-1}{2}}\nonumber\\
&=&2\Gamma\left(-\frac{D+1}{2}\right)\zeta\left(-\frac{D-1}{2},\frac{a^2\mu^2}{4\pi^2}\right)a^{1-D}
+\frac{\sqrt{\pi}a^{D+1}}{(ah)^D}\Gamma\left(-\frac D2\right)\zeta\left(-\frac D2,\frac{\mu^2h^2}{4\pi^2}\right)\nonumber\\
&+&\frac{8\pi^{\frac{1-D}{2}}}{a^{\frac D2-1}h^{\frac D2}}\sum_{n_1,n_2=1}^{\infty}\left(\frac{\sqrt{n_2^2+\left(\frac{\mu}{2\pi}\right)^2h^2}}{n_1}\right)^{\frac D2}K_{\frac D2}\left(\frac{2\pi an_1}{h}\sqrt{n_2^2+\left(\frac{\mu}{2\pi}\right)^2h^2}\right).
\label{eq24}
\end{eqnarray}
From the relation between the Epstein-Hurwitz $\zeta$ function and
the Riemann $\zeta$ function $\zeta(s,0)=\zeta(2s)$ and
Eq.(\ref{eq19}), we obtain again the massless expression
 when $\mu\rightarrow 0$.

Using Eqs.(\ref{eq18}) and (\ref{eq24}), we have the global energy between the separation $a$ from Eq.(16) as
\begin{eqnarray}
E^D&=&\frac 12 \pi^{\frac{D-1}{2}}\Gamma\left(\frac{1-D}{2}\right)\left(\frac{\mu}{2\pi}\right)^{D-1}-\frac 12\pi^{\frac {D+1}{2}}\Gamma\left(-\frac{D+1}{2}\right)\left(\frac{\mu}{2\pi}\right)^{D+1}ah\nonumber\\
&-&2\left(\frac{\mu a}{2\pi}\right)^{\frac D2}a^{1-D}\sum_{n_2=1}^{\infty}n_2^{-\frac D2}K_{\frac D2}(n_2\mu a)\nonumber\\
&-&2\left(\frac{\mu h}{2\pi}\right)^{\frac {D+1}2}h^{-D}a\sum_{n_2=1}^{\infty}n_2^{-\frac {D+1}2}K_{\frac {D+1}2}(n_2\mu h)\nonumber\\
&-&4a^{1-\frac D2}h^{-\frac D2}\sum_{n_1,n_2=1}^{\infty}\left(\frac{\sqrt{n_2^2+\left(\frac{\mu}{2\pi}\right)^2h^2}}{n_1}\right)^{\frac D2}K_{\frac D2}\left(\frac{2\pi an_1}{h}\sqrt{n_2^2+\left(\frac{\mu}{2\pi}\right)^2h^2}\right).\nonumber\\
\label{eq25}
\end{eqnarray}

Ambj{\o}rn and Wolfram \cite{22} have treated the case of Dirichlet
boundary condition for a massive scalar field. Following their
methodology, we get the finite physically relevant energy. The first
term in Eq.(\ref{eq25}) gives a contribution to the total energy
independent of $a$ and $h$, therefore it can be dropped. The second
term in Eq. (\ref{eq25}) corresponds to a constant energy density
which can be canceled by addition of a constant to the Hamiltonian
density. Therefore, the finite physically relevant energy is
\begin{eqnarray}
E^D=&-&2\left(\frac{\mu a}{2\pi}\right)^{\frac D2}a^{1-D}\sum_{n_2=1}^{\infty}n_2^{-\frac D2}K_{\frac D2}(n_2\mu a)\nonumber\\
&-&2\left(\frac{\mu h}{2\pi}\right)^{\frac {D+1}2}h^{-D}a\sum_{n_2=1}^{\infty}n_2^{-\frac {D+1}2}K_{\frac {D+1}2}(n_2\mu h)\nonumber\\
&-&4a^{1-\frac D2}h^{-\frac D2}\sum_{n_1,n_2=1}^{\infty}\left(\frac{\sqrt{n_2^2+\left(\frac{\mu}{2\pi}\right)^2h^2}}{n_1}\right)^{\frac D2}K_{\frac D2}\left(\frac{2\pi an_1}{h}\sqrt{n_2^2+\left(\frac{\mu}{2\pi}\right)^2h^2}\right).\nonumber\\
\label{eq26}
\end{eqnarray}

Using the expressions of the modified Bessel function
\begin{equation}
K_{j+\frac
12}(z)=\sqrt{\frac{\pi}{2z}}e^{-z}\sum_{k=0}^{j}\frac{(j+k)!}{k!(j-k)!(2z)^{k}},\label{eq27}
\end{equation}
and
\begin{eqnarray}
K_{j}(z)=& &\frac
12\sum_{k=0}^{j-1}(-1)^k\frac{(j-k-1)!}{k!\left(\frac z
2\right)^{j-2k}}+(-1)^{j+1}\sum_{k=0}^{\infty}\frac{\left(\frac z
2\right)^{j+2k}}{k!(j+k)!}\nonumber\\
&\times&\left[\ln\frac 32-\frac 12\psi(k+1)-\frac
12\psi(j+k+1)\right],\label{eq28}
\end{eqnarray}
where $\psi(k+1)=-c+(1+\frac 12+\cdots+\frac 1k)$ and $c$ is the
Euler's constant, we can calculate the Casimir energy for massive
case. Especially, we have the physical Casimir energy for small mass
$\mu$ from Eqs.(\ref{eq26})-(\ref{eq28}) up to $\mu^2$ order,
\begin{eqnarray}
E^D=& &E_{\mu=0}^D +\frac 14a^{1-D}\pi^{-\frac D
2}\Gamma\left(\frac{D-2}2\right)\zeta(D-2)(1-\delta_D^3)(\mu
a)^2\nonumber\\
&+&\frac 1 4h^{-D}a\pi^{-\frac{D+1}2}\Gamma\left(\frac
{D-1}2\right)\zeta(D-1)(1-\delta_D^2)(\mu h)^2\nonumber\\
&+&\frac 1{\pi}a^{\frac
{4-D}2}h^{-\frac{D+2}2}\sum_{n_1,n_2=1}^{\infty}\left(\frac{n_2}{n_1}\right)^{\frac{D-2}2}K_{\frac
{D-2}2}\left(\frac{2\pi an_1n_2}{h}\right)(\mu h)^2,\label{eq29}
\end{eqnarray}
where $\delta_D^2$ and $\delta_D^3$ are the Kronecker delta. When
$\mu a\gg 1$ and $\mu h\gg 1$, the approximate energy is
\begin{equation}
E^D\sim -3a^{1-D}\left(\frac{\mu a}{2\pi}\right)^{\frac
{D-1}2}e^{-\mu a}-ah^{-D}\left(\frac{\mu h}{2\pi}\right)^{\frac
{D}2}e^{-\mu h}.\label{eq30}
\end{equation}

In the $D=3$ case, Eq.(\ref{eq29}) is reduced to
\begin{equation}
E=E_{\mu=0} +\left[\frac a{24h}+\frac 1{\pi}a^{\frac 12}h^{-\frac
12}\sum_{n_1,n_2=1}^{\infty}\left(\frac{n_2}{n_1}\right)^{\frac
12}K_{\frac 12}\left(\frac{2\pi an_1n_2}{h}\right)\right]\mu
^2.\label{eq31}
\end{equation}
The pressure on the $x^1$ direction is
\begin{equation}
P_a=P_{a,\mu=0} +\left[\frac 1{24h}-2a^{\frac 12}h^{-\frac
32}\sum_{n_1,n_2=1}^{\infty}n_2^{\frac 32}n_1^{\frac 12}K_{\frac
12}\left(\frac{2\pi an_1n_2}{h}\right)\right]\mu ^2.\label{eq32}
\end{equation}
By the numerical calculation, we plot $\Delta P_a=P_a-P_{a,\mu=0}$
in Fig. 3 for $h=4$ and we find $\Delta P_a=0$ appears at
$a=r_{crit}^{\prime}h$, where $r_{crit}^{\prime}=0.523522\cdots$.
That is, the pressure is independent of the mass at
$r=r_{crit}^{\prime}$. When $a<r_{crit}^{\prime}h$, $\Delta P_a<0$
and when $a>r_{crit}^{\prime}h$, $\Delta P_a>0$. The larger the mass
$\mu$ is, the more deviation is.
\begin{figure}[ph]
\centerline{\psfig{file=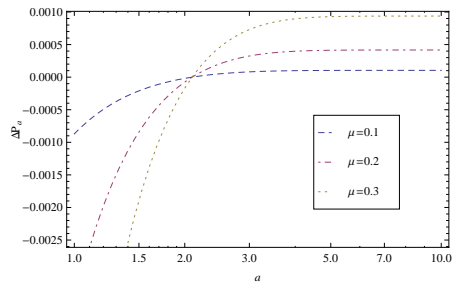,width=2.5in}} \caption{The
difference of the pressure of massive field and massless field on
the $x^1$ direction $\Delta P_a=P_a-P_{a,\mu=0}$ for $D=3,h=4$ and
$\mu=0.1,0.2,0.3$, respectively. It is clearly seen that $\Delta
P_a=0$ appears at $a=r_{crit}^{\prime}h$, where
$r_{crit}^{\prime}=0.523522\cdots$. When $a<r_{crit}^{\prime}h$,
$\Delta P_a<0$ and when $a>r_{crit}^{\prime}h$, $\Delta P_a>0$. The
larger the mass $\mu$ is, the more deviation of $\Delta P_a$ from
$\Delta P_a=0$ is.}
\end{figure}
\section{Conclusions }

In this work we use the Chowla-Selberg formula and its
generalization to calculate the Casimir effect with a helix boundary
condition in ($D+1$)-dimensional spacetime. We obtain the exact
results of the Casimir energy density and pressure for any $D$ for
both massless and massive scalar fields.

In this work, we present for the first time (as far as we know) a
topology of flat ($D+1$)-dimensional spacetime. This topology causes
the helix boundary condition for a Hermitian scalar field. With the
new boundary condition, the spectrum of the field will have new
feature, which will lead to the observable effect.

The main conclusions of this work are:

$\bullet$ Once topology of spacetime is fixed, the ratio of the sizes of the helix
will be a decisive factor. In our case, the negative pressure will
become positive with the ratio $r$ of the lengths increasing. This
phenomenon of quantum physics is similar to that in the rectangular
cavity with Dirichlet condition\cite{8,9}.

$\bullet$ Both in massless and massive cases, there are critical
values $r_{crit}(\mu)$. When $a=r_{crit}h$ or $h=r_{crit}a$, the
pressure $P_a$ or $P_h$ vanishes. In the $\mu=0$ case,
$r_{crit}(0)=1.52007606\cdots$ for $D=3$.

$\bullet$ For a massive field, when the mass $\mu$ tends to zero,
the pressure approaches the result of the pressure in massless case
and when the mass $\mu\gg 1$, the pressure for a massive field goes
to zero. There is another critical value $r_{crit}^{\prime}$ and the
pressure is independent of the mass $\mu$ at $r=r_{crit}^{\prime}$.
In the $D=3$ case, $r_{crit}^{\prime}=0.523522\cdots$.

$\bullet$ The summation formulae in the number theory are very
useful such as the Chowla-Selberg formula and its generalization for
the quantum physics.

Further, the numerical calculation shows that there is a $Z_2$
symmetry of $a\leftrightarrow h$. It is not surprised because the
boundary conditions (\ref{eq6}) and (\ref{eq7}) are equivalent to
(\ref{eq6}) and (\ref{eq8}) . Therefore, both $P_a$ and $P_h$ have
the same characteristics.

\section*{Acknowledgments}
This work is supported by National Nature Science Foundation of
China under Grant Nos.10671128 and 11047138, the Key Project of
Chinese Ministry of Education.(No211059), Innovation Program of
Shanghai Municipal Education Commission(11zz123), National Education
Foundation of China Grant No. 2009312711004 and Shanghai Natural
Science Foundation , grant No. 10ZR1422000.


\begin{thebibliography}{999}

\bibitem{1}
M. Bordag, G. L. klimchitskaya, U. Mohideen and V. M. Mostepanenko,
{\it Advances in the Casimir Effect}, Oxford: Oxford University
Press, 2009.
\bibitem{2}
C. J. Feng and X. Z. Li, \textit{Phys. Lett. B} {\bf 691},
167(2010).
\bibitem{3}
X. H. Zhai, X. Z.Li and C. J. Feng, \textit{Mod. Phys. Lett. A}
\textbf{26}, 669(2011).
\bibitem{4} J. R. Munkres, \textit{Elements of Algebric Topology},
Addison-Wesley Publishing Company, Amdterdam, 1984.

\bibitem{5}
  E.~Elizalde, S.~D.~Odintsov, A.~Romeo, A.~A.~Bytsenko and S.~Zerbini, \textit{Zeta Regularization Techniques with
  Applications}, World Scientific, Singapore, 1993.

\bibitem{6}
  X.~Z.~Li, X.~Shi and J.~Z.~Zhang,
  \textit{Phys.\ Rev.\  D} {\bf 44}, 560(1991).

\bibitem{7} I.~H.~Brevik and E. Elizalde,
 \textit{ Phys.\ Rev.\  D} {\bf 49}, 5319(1994).

\bibitem{8}
  X.~Z.~Li, H.~B.~Cheng, J.~M.~Li and X.~H.~Zhai,
  \textit{Phys.\ Rev.\  D} {\bf 56}, 2155 (1997).

\bibitem {9} X.~Z.~Li and X.~H.~Zhai,
  \textit{J.\ Phys.\ A} {\bf34}, 11053(2001).


\bibitem{10} S. C. Lim and L. P. Teo, \textit{J. Phys. A} {\bf 40}, 11645 (2007).

\bibitem{11} L. P. Teo, \textit{Phys. Rev. D} {\bf 82}, 105002 (2010).

\bibitem{12}
  X.~Shi and X.~Z.~Li,
  \textit{Class.\ Quant.\ Grav.}\  {\bf 8}, 75(1991).

\bibitem{13}
    R.~M.~Cavalcanti,
  \textit{Phys.\ Rev.\  D} {\bf 69}, 065015(2004).

\bibitem{14}
  M.~P.~Hertzberg, R.~L.~Jaffe, M.~Kardar and A.~Scardicchio,
 \textit{ Phys.\ Rev.\ Lett.}\  {\bf 95}, 250402(2005).

\bibitem{15}
  X.~H.~Zhai and X.~Z.~Li,
  \textit{Phys.\ Rev.\  D }{\bf 76}, 047704(2007).

  \bibitem{16} X.~H.~Zhai, Y.~Y.~Zhang and X.~Z.~Li,
    \textit{Mod.\ Phys.\ Lett.\  A }{\bf 24}, 393(2009).

\bibitem{17} S. C. Lim and L. P. Teo, \textit{Annals Phys. }\textbf{324}, 1676(2009).

\bibitem{18} C. H. Eab, S. C. Lim and L. P. Teo, \textit{J. Math. Phys.} {\bf 48}, 082301 (2007).

\bibitem{19} S. C. Lim and L. P. Teo, \textit{Phys. Lett. B }{\bf 679}, 130 (2009).

\bibitem{20} E. Elizalde, \textit{J. Phys. A }{\bf 27}, 3775 (1994).

\bibitem{21}E. K. Abalo, K. A. Milton and L. Kaplan, \textit{Phys. Rev. D} \textbf{82}, 125007 (2010).

\bibitem{22} J. Ambj{\o}n and S. Wolfram, \textit{Annals Phys.} {\bf 147}, 1 (1983).





\end{thebibliography}
\end{document}